\documentclass{article}

\usepackage[utf8]{inputenc} 
\usepackage[T1]{fontenc}    
\usepackage{hyperref}       
\hypersetup{colorlinks, citecolor=blue, linkcolor=red, urlcolor=black}
\usepackage{url}            
\usepackage{booktabs}       
\usepackage{amsfonts}       
\usepackage{amssymb}
\usepackage{nicefrac}       
\usepackage{microtype}      
\usepackage{graphicx}       
\usepackage{xcolor}         
\usepackage{siunitx}
\usepackage{amsmath}

\usepackage[draft]{changes}

\usepackage[left=1.2in, right=1.2in, top=1in, bottom=1in]{geometry}
\usepackage[style=nature, backend=biber]{biblatex} 
\usepackage[labelfont=bf,textfont=md]{caption}
\usepackage{algorithm, algpseudocode}
\usepackage{authblk}
\usepackage{chemformula}

\title{\textbf{Grand canonical generative diffusion model for \\crystalline phases and grain boundaries}}

\author[1]{Bo Lei}
\author[1]{Enze Chen}
\author[1]{Hyuna Kwon}
\author[1]{Tim Hsu}
\author[1]{Babak Sadigh}
\author[1]{Vincenzo Lordi}
\author[1]{Timofey Frolov}
\author[1]{Fei Zhou\thanks{zhou6@llnl.gov}}

\affil[1]{Lawrence Livermore National Laboratory, Livermore, CA 94551, USA}
\date{}
\begin{document}
\maketitle
\begin{abstract}
The diffusion model has emerged as a powerful tool for generating atomic structures for materials science. 
This work calls attention to the deficiency of current particle-based diffusion models, which represent atoms as a point cloud, in generating even the simplest ordered crystalline structures. The problem is attributed to particles being trapped in local minima during the score-driven simulated annealing of the diffusion process, similar to the physical process of force-driven simulated annealing. We develop a solution, the grand canonical diffusion model, which adopts an alternative voxel-based representation with continuous rather than fixed number of particles. The method is applied towards generation of several common crystalline phases as well as the technologically important and challenging problem of grain boundary structures.
\end{abstract}

\begin{figure}[!ht]
    \centering
    \includegraphics[width=0.95\linewidth]{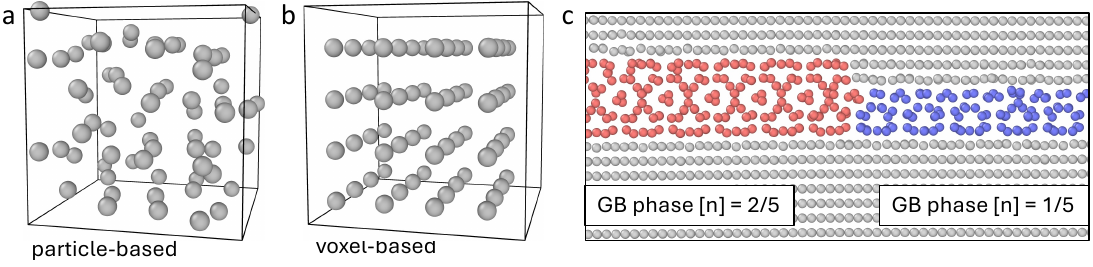}
    \caption{Comparison of diffusion models for simple cubic structures: (a) particle-based generator with fixed number of particles, and (b) voxel-based grand canonical generator with continuously adjustable number of particles. (c) 
    Grain boundary phase transition in tungsten $\Sigma5(100)$ twist boundary. The two GB phases with different atomic densities $[n]=1/5$, $2/5$ are shown in blue and red, respectively. The bulk atoms of the misoriented upper and lower crystals are shown in gray.}
    \label{fig:upfront-DPM-comparison}
\end{figure}

\section{Introduction}

The ability to traverse and sample from the space of atomic structures is a ubiquitous and indispensable component of computational materials science. For example, to accelerate materials discovery, inverse design of materials starts with target material properties and seeks compositions and structures that satisfy the specifications \cite{sanchez2018inverse}. High-throughput materials screening and data mining pipelines need to explore novel structures beyond known datasets to broaden their scope of search \cite{Fischer2006NM,Jain2013APLM}. Mesoscale or coarse-grained materials modeling, e.g. on materials strength, increasingly resort to atomistic simulations such as molecular dynamics on structures that correspond to a coarse-level configuration for improved parametrization and verification \cite{Bertin2023}. Property-prediction efforts such as interatomic potential development need to sample new structures outside the training set to ensure transferability of the trained models \cite{Deringer2019AM,Mishin2021AM}. While important and useful, this \textit{inverse} mapping from property to structure can be more challenging than the straightforward \textit{forward} map or structure-property relationships. The inverse map is often one-to-many, as there can be many configurations in the high-dimensional structure space that satisfy the desired criterion. It may even be ill defined if there is no such structure that exactly meets the goal.
%
Well-known traditional approaches to generate atomic structures include 
molecular dynamics and Monte-Carlo to sample from a statistical distribution (often Boltzmann), and structure optimization/prediction methods, a special case for sampling low energy structures, such as simulated annealing \cite{Pannetier1990N-simulated-annealing}, genetic algorithm \cite{Deaven1995PRL-genetic,Oganov2011ACR}, and basin hopping \cite{Wales1997JPCA}.  They require knowledge of a potential energy surface to drive the sampling or optimization and are typically computationally expensive for large systems.

The rapid ascent of generative artificial intelligence techniques, especially probabilistic methods such as normalizing flow \cite{Altieri2015} and diffusion probabilistic model (DPM) \cite{Sohl-Dickstein2015-DPM, Ho2020-DDPM, Song2020-unified}, has brought fresh tools to the long-standing challenge of structure generation \cite{noh2019inverse,yao2021inverse,kim2020inverse,noh2020machine}.  
Among its numerous achievements \cite{Yang2022-dpm-review}, the diffusion model has been applied to generating materials such crystal and molecular structures. While the initial efforts have been focused on generating molecules, especially organic molecules \cite{hoogeboom2022equivariant, jing2022torsional, Xu2022-dpm-molecule,Wu2022NC, weiss2023guided, Wu2022NC, Duan2023NCS, Ingraham2023N, Watson2023N, Lu2023, Corso2022-DiffDock, Jing2023-EigenFold, Hsu2024JCTC-SD}, recently there has been an increasing number of works targeting crystalline solids \cite{Xie2022-cdvae, jiao2024crystal, zeni2023mattergen, yang2023scalable, pakornchote2024diffusion}. 
DPM formulates the inverse map as a well-defined conditional probabilistic problem: to generate/sample candidate structures conditioned on a given criterion. In the training phase, it learns the training dataset's probability distribution, more specifically the score $\partial_{\boldsymbol{x}} \log P(\boldsymbol{x})$ or the derivative of log probability with respect to (high dimensional) data $\boldsymbol{x}$. This is done through score matching, typically denoising score matching \cite{Vincent2011scorematching, Song2019NeurIPS-Generative}. In the inference phase, one integrates a diffusion equation starting from random initial points in the data space under the driving force of the learned score towards denoised outcomes according to the target distribution. This equation of motion can be either a deterministic ordinary differential equation (ODE) or a stochastic differential equation (SDE) formulated as a Langevin equation \cite{Song2020-unified}. When it comes to structure generation with real atomic coordinates $\boldsymbol{x}$, the above process looks conspicuously similar to the well-known \textit{simulated annealing} (SA) algorithm for simulating physical systems driven by atomic forces (rather than score) \cite{Pannetier1990N-simulated-annealing}. The similarity and connection to statistical mechanics were of course explicitly called out and cited as an aspiration to DPM \cite{Song2019NeurIPS-Generative,Song2020-unified, Dockhorn2022ICLR-Score}.

\begin{figure}[!ht]
    \centering
    \includegraphics[width=0.75\linewidth]{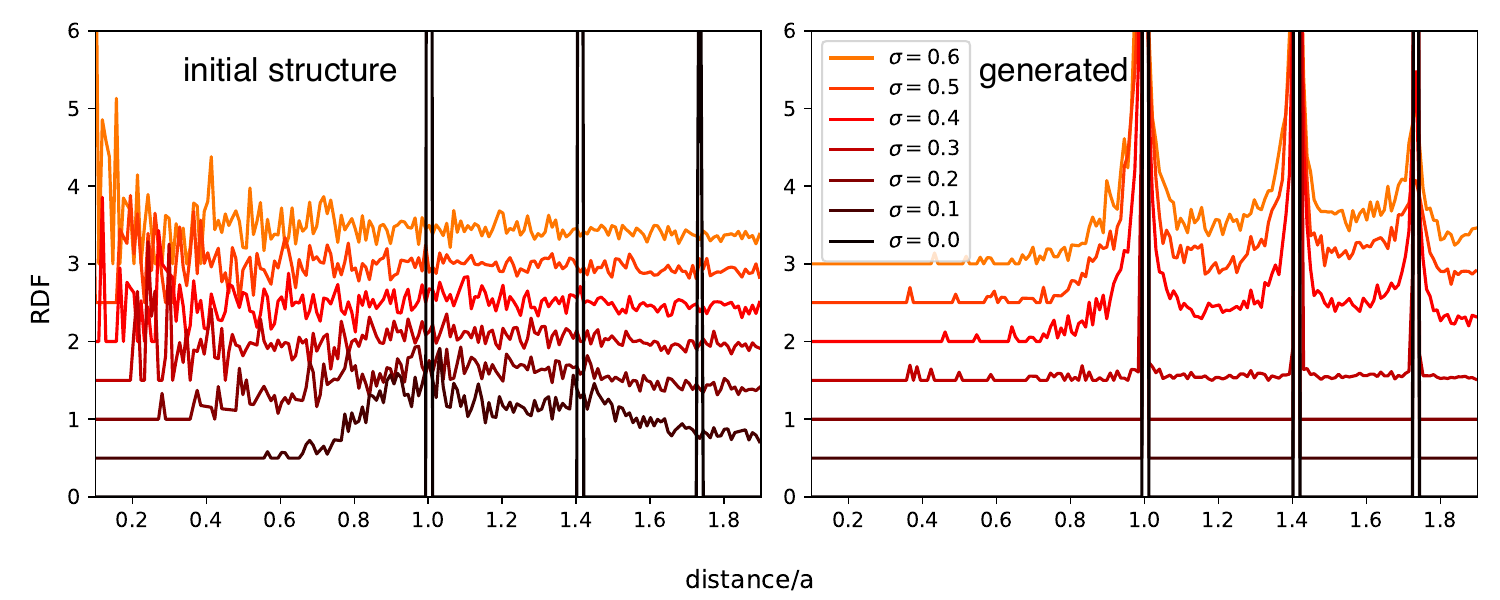}
    \caption{Effects of different levels of randomness $\sigma$ in initial structures on particle diffusion models. $\sigma=0 \ (\gtrapprox 0.5)$ means unperturbed (random) initial structure. Radial distribution functions of simple cubic structure (lattice constant 1) averaged over ten $4\times 4\times 4$ supercells are shown.  Left: initial structures of ideal SC+Gaussian displacements $\mathcal{N}(0,\sigma^2)$; right: generated with diffusion.}
    \label{fig:sc-rdf-perturb}
\end{figure}
Given SA's crucial role in the diffusion model, it is interesting to investigate whether its limitations in physical simulation are also present in DPM. A well-known issue is that SA can get trapped in local minima and miss the long-range ordered crystalline phase as the correct global minimum, at least not without introducing defects in the sampled structures \cite{Bollweg1997}. Indeed, we have observed the same problem in particle-based DPM. The failure was recently noted for crystalline diamond in Ref.~\cite{Kwon2023-Spectroscopy}. Fig.~\ref{fig:upfront-DPM-comparison}a shows a particle-based DPM trained on a simple cubic (SC) crystal structure fails to generate SC starting from random uniform positions, the asymptotic distribution of indistinguishable particle with large i.i.d. displacements $\sim \mathcal{N}(0,\sigma^2)$ according to the variance exploding noise schedule of DPM. 
In contrast, our grand canonical diffusion model was able to generate crystalline structures without any problem (details in Results), as shown in Fig.~\ref{fig:upfront-DPM-comparison}b.
Fig.~\ref{fig:sc-rdf-perturb} quantitatively illustrates the effects of initial randomness on SC generation with this particle-based DPM. Initialized with up to 20\% Gaussian displacements added to SC (left), it can successfully generate ordered SC evidenced by sharp radial distribution function (RDF) peaks (right). At 30\% initial noise, structural defects emerge. With 50\% or more, the SC structure is ``forgotten'', the RDF is featureless and the initial structures are essentially random uniform (left), leading to substantial RDF broadening in the generated structures (right). In short, a particle DPM can generate ordered SC only when the cubic topology is present in the initial structure. Otherwise, it cannot tolerate initial randomness and gets trapped in local minima.
%
\paragraph{Grain Boundary}
Crystalline phases, if already known and available as a training set, are relatively simple. 
More interesting and technologically relevant are crystalline structures with defects. 
Most engineering materials are composed of many crystallites or grains separated by interfaces called grain boundaries (GBs), which are the primary goal of this paper (see Fig.~\ref{fig:upfront-DPM-comparison}c for an example).
These interfaces, inherited from synthesis and processing, greatly influence many important properties of polycrystalline materials including mechanical, electrical, thermal, corrosion resistance, diffusion, and optical properties. 
Recently, it has been recognized that even though GBs are effectively the disruptions of perfect crystalline order, they have a well-defined and ordered structure~\cite{frolov_2013_gbphase, vonalfthan_2006}.
The five microscopic degrees of freedom associated with the orientation of the abutting crystals make the space of possible GB structures extraordinarily vast~\cite{sutton_1995}. 
However, even for a fixed crystal orientation and the GB plane, the interface structure can also exist in multiple states or GB phases characterized by the different atomic structures of the boundary. 
First-order GB phase transitions have been directly observed by high-resolution electron microscopy~\cite{meiners_2020} and atomistic simulations~\cite{frolov_2013_gbphase}.
Upon a transition, the boundary's structure and properties change in a discontinuous manner which in turn leads to abrupt changes in properties of polycrystalline materials. 
Computational tools are needed that can efficiently predict the structure and properties of GBs to discover how to manipulate them with chemical composition, temperature, or pressure.

Even in the simplest elemental metals, GB structure prediction can be extremely challenging and require advanced grand canonical optimization~\cite{zhu_2018}. 
GBs of the same orientation can exist in many metastable states of different relative translations and different number of atoms separated by large energy barriers, which prevents the interface from finding its ground state on the short timescale of simulations or even experiments. 
It is generally insufficient to simply take two perfect half crystals and join them together to obtain the correct ground state. 
Instead, advanced structure sampling methods such as genetic algorithms~\cite{zhu_2018} and simulated annealing~\cite{vonalfthan_2006}  have been applied to predict GBs by inserting or removing atoms from the GB core followed by atomic position optimization.
These computationally intense algorithms require systematic exploration of a complex structural space, and their efficiency can be improved tremendously by proposing better initial candidate structures and proposing intelligent mutation moves aimed at improving the current configurations~\cite{zhu_2018}. Such searches 
%
explore different metastable states and often generate hundreds or thousands of 
candidate structures before they find the ground state. 
This generated rich information about the different GB atomic environments is usually not utilized in the follow-up searches, and for every new boundary, the search is performed from scratch. 
It is well known, however, that many different boundaries are not completely different since they may be composed of the same dislocations with the same dislocation core structure~\cite{frolov_2013_gbphase, zhu_2018, frolov_2018_grain}. 
Therefore, it is desirable to develop AI algorithms that could learn the GB structural space to accelerate structure searches and identify hidden relations for GBs.
Given the aforementioned difficulty of generating simple crystalline phases, GBs with both ordered and disordered atoms are expected to be challenging.

The main contributions of this paper are
\begin{itemize}
    \item To call attention to the deficiency of current particle-based diffusion models in generating even the simplest ordered crystalline structures, and to explain the reason for the failure.
    \item To provide a solution---the grand canonical diffusion model---with an alternative voxel-based representation and variable number of particles.
    \item To apply the model for generation of grain boundary structures, which are important in many materials problems.
\end{itemize}


\section{Method}

\subsection{Related work: Voxel-based generative model for atomic structure}
Pinheiro \textit{et al} designed a voxel-based model for molecule generation \cite{Pinheiro2023NeurIPS-3D}. There are a few important distinctions. First, ref.~\cite{Pinheiro2023NeurIPS-3D} did not use a diffusion model. Second, Pinheiro \textit{et al} employed a discrete grid and involved an extra step to convert the grid to atomic coordinates. Here, we have smeared the positions, which leads to the ability to sample continuous coordinates, and we focus on solid materials.
\begin{figure}[htp]
\centering
\includegraphics[width=0.8 \linewidth]{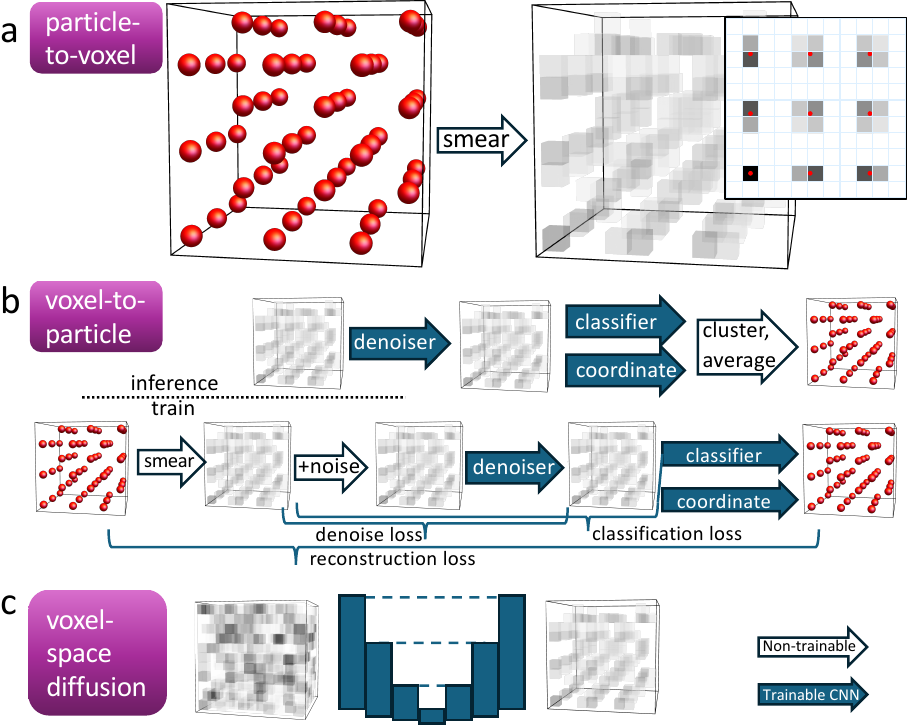} 
\caption{Grand canonical diffusion model with voxel representation. (a) Particle to voxel encoder by smearing using linear weights and no learnable parameters. For clarity a 2D slice is also shown with red dots designating particle positions. (b) Learnable voxel to particle coordinate decoder. Both the training and inference stages are shown. 
(c) Voxel-space diffusion using a convolutional U-Net. 
} \label{fig:model} 
\end{figure}


\subsection{Particle-based DPM and GNN score model}
We adopted a baseline particle-based GNN diffusion model from Ref.~\cite{Kwon2023-Spectroscopy}. Throughout this work we are concerned with single-element materials, and we will regard particle/atom type encoding as fixed and optional. The variance-exploding noise schedule was used, under which perturbed atomic coordinates approach a random uniform distribution. This enables one to start the denoising trajectory from a random structure sampled from a uniform distribution over the coordinate space. Attempts were made to improve structure sampling and avoid local minima, including testing both the ODE and the SDE, increasing the integration steps to 1000, and the restart sampling algorithm \cite{xu2023restart}, but local minima persisted.  More details can be found in the Appendix.

\subsection{Grand canonical voxel diffusion model} \label{sec:GC}
\paragraph{Particle-to-voxel encoder.}
An encoder $\text{EN}$ converts an input structure or point cloud  $\boldsymbol{R}=\{\boldsymbol{r}\}$ into a voxel array $C=\text{EN}(\boldsymbol{R})$ by smearing the coordinates of each particle (Fig~\ref{fig:model}a). One assumes $N$ particles, with coordinate $r_{a \alpha}$ for particle (atom) $a$ in direction $\alpha=1,2,3$, and  spatial partitions $\{V_I \}$ with each partition $V_{I}$ representing a region of the simulation cell $\Omega = \bigcup_I V_I$ without overlap ($V_I \cap V_J = \empty$ for $I \neq J$). Here, each partition $ V_I = V_{ijk} = \{\boldsymbol{r}| \boldsymbol{r} \in V, (i-\frac12)L \leq r_{1} < (i+\frac12)L, \dots  \}$ is a cube of linear size $L$ centered at $(iL,jL,kL)$. Together they form a regular grid. Each particle $a$ is smeared from a point or delta distribution into a uniform distribution $U(\boldsymbol{r}-\boldsymbol{r}_a)$ within a size-$L$ cube, and contribute $ P(a, I)  $ to partition $V_I$. $P(a, I)$ is the proportion or relative volume of smeared ``cubic'' particle $a$ within $V_I$,
\begin{align} \label{eq:proportion}
  P(a, I) &= \int_{V_I} U(\boldsymbol{r}-\boldsymbol{r}_a) d\boldsymbol{r} = \prod_{\alpha=1}^{3} P(r_{a\alpha}, I_\alpha), \\
  P(r, i) &= \begin{cases}
    1-|r- iL|/L,& \text{if }  |r- iL|/L < 1\\
    0,              & \text{otherwise}
            \end{cases} \nonumber
\end{align}
The above formulation is trivially generalized for periodic boundary conditions, which are often adopted for crystal structures. If a particle is exactly at the center of a partition, it contributes wholly 1 to that partition, as shown in the lower left corner of 2D inset in Fig~\ref{fig:model}a. More generally, each particle is smeared into up to 8 (4) neighbor voxels in 3D (pixels in 2D). The total voxel array $C$ is summed over all particles
\begin{align} \label{eq:voxel}
 C_I =  \sum_{a} P(a, I).
\end{align}
To avoid associating any voxel with more than one particle and obscuring the voxel-based generator and the decoder, $L$ is preferably a small fraction of the shortest bond distance. There is a compromise in the choice of $L$: smaller $L$ means less smearing and more precise locations, but more voxels and larger memory footprint, which becomes a problem for large 3D cells. 

Our encoder design, with a grid-size hyperparameter $L$ and no learnable parameters,  encodes the particle coordinates continuously into the voxel array $C$, as is evident in Eqs.~(\ref{eq:proportion},\ref{eq:voxel}). This allows for the voxel-space generator to learn on a continuous voxel representation without discontinuity. 

Additionally, a data-augmentation step is adopted to enforce translational equivariance of our approach. By construction the encoder is not exactly translationally equivariant: if all atoms are translated by \textit{integer} multiplets of $L$, then the array $C$ is the same after translation. 
Otherwise, translation by fractional multiplets of $L$ does modify $C$, as seen in the 2D inset in Fig.~\ref{fig:model}a. In the training step random translation vectors  from a uniform distribution $U(0,L)$ are sampled. The same data-augmentation step is adopted throughout the paper to train the decoder and the voxel diffusion model.

\paragraph{Voxel-to-particle decoder}
Since the coordinates were continuously encoded into voxel values, a dedicated decoder $\text{DE}$ based on a 3D convolutional neural network (CNN) was trained to map voxel arrays into coordinates (Fig.~\ref{fig:model}b). It consists of three modules, each being a single-layer ConvNext model \cite{Liu2022-convnext} with $3^3$ filters and 64 hidden features: a denoiser $\text{DE}_\text{den}$ to reduce noise in the input voxel array, a binary classifier $\text{DE}_\text{cls}$ to predict whether a voxel is associated with or occupied by any particle, and a relative coordinate regression model $\text{DE}_\text{cor}$ that predict the cartesian coordinate of any associated particle. The decoder was trained with three corresponding loss functions (lower part of Fig.~\ref{fig:model}b). First, we introduce a voxel value denoising loss inspired by denoising score matching \cite{Vincent2011scorematching} 
\begin{align}
    L_\text{den} &=   \left\| \text{EN}(\boldsymbol{R}) - \text{DE}_\text{den}(\text{EN}(\boldsymbol{R})+\sigma)  \right\|_2^2 ,
\end{align}
where structure $\boldsymbol{R}$ is sampled from the same training atomic structures as the generator model, and $\sigma \sim  \mathcal{N}(0,\sigma_{\max}^2)$ is i.i.d. Gaussian noise of magnitude $\sigma_{\max}$. In the decoder training stage, random noises are added to the encoded (smeared) voxel configuration, and the denoising module $\text{DE}_\text{den}$ and loss $L_\text{den}$ help with the quality and stability of decoding.
Second, a classification loss employing binary cross entropy
\begin{align}
    L_\text{cls} &= \text{BCE}(\text{EN}(\boldsymbol{R})>0, \text{DE}_\text{cls}(\text{DE}_\text{den}(\text{EN}(\boldsymbol{R})+\sigma)) )
\end{align}
helps identify the voxels with particle occupancy in them, i.e. non-vanishing and substantial overlap $P(a,I)$. The classifier was trained against the ground truth logical label $\text{EN}(\boldsymbol{R})> p_{\text{thre}}$. Up to 8 voxels can be partially occupied by a particle and the occupancies are 1/8 in the most even case, so the threshold $p_{\text{thre}}$ for substantial overlap was deliberately chosen to a small 0.05.
Thirdly, a coordinate reconstruction loss
\begin{align}
    L_\text{cor} &=  \left\| \text{int}(\text{DE}_\text{cls}(\text{DE}_\text{den}(\text{EN}(\boldsymbol{R})+\sigma))) ( \Delta \boldsymbol{r} - \text{DE}_\text{cor}(\text{DE}_\text{den}(\text{EN}(\boldsymbol{R})+\sigma))) \right\|_2^2
\end{align}
recovers the coordinate of the potential particle relative to the partition's center. $\Delta \boldsymbol{r}$ is the relative displacement between the associated particle and center of a partition. Partitions without a predicted overlapping particle are disregarded in this loss by $\text{int}$, which converts true/false to 1/0. 
The total loss is
\begin{align}
    L =\mathbb{E}_{\boldsymbol{R}, \sigma, \boldsymbol{t}} \left[ (L_\text{den} +  L_\text{cls} + L_\text{cor} ) |_{\boldsymbol{R}+\boldsymbol{t},\sigma} \right],
\end{align}
where vector $\boldsymbol{t}$ is sampled from uniform  $t_\alpha \sim U(0, L)$ to translate all atoms in each structure $\boldsymbol{R}$ such that the smeared voxels are approximately equivariant under translation.

For inference (upper part of Fig.~\ref{fig:model}b), the input voxel array $C$ is first denoised into $C'=\text{DE}_\text{den}(C)$, partitions with particles are  identified by the classifier $\text{DE}_\text{cls}(C')$, and then coordinates of a particle is found for each identified partition $ijk$ as $(iL,jL,kL)+ \text{DE}_\text{cor}(C') $. This algorithm tend to conservatively return more than one closely distributed coordinates for each atom due to smearing. Finally, duplicates were removed by a refinement process, which finds the cluster of coordinates based on the euclidean distance between pairs and return the averaged coordinate for each cluster. A visualization of the decoder's training and inference can be found in Fig.~\ref{fig:model}b.

\paragraph{Voxel-space diffusion model.}
With the encoder and decoder to map between particles and voxels, a voxel-space diffusion model was trained that adopts a convolutional U-Net, largely following Ref.~\cite{Saharia2021-Palette}. 
To treat 3D voxel data,  2D operators such as convolution, pooling, normalization, up-sampling and down-sampling were replaced with their 3D counterparts. Periodic boundary conditions were adopted for crystalline phases, while special hybrid boundary conditions were used for GBs, which were periodic in the $xy$ directions but capped with vacuum in both ends of the $z$ direction. Therefore we applied periodic padding in $xy$ and zero padding in $z$. The SiLU activation function \cite{Hendrycks2016-GELU} and variance-preserving noise schedule were adopted.
We scaled voxels $C\leftarrow C/s$ by $s=0.02$, which is approximately the standard deviation of the smeared voxel values, most of which are zero. The \textit{solver3} variant of DPM-Solver \cite{Lu2022-DPM-solver} was employed 
with 60 function evaluations.

\paragraph{Model training.}\label{sec:training} All learnable networks were trained with an AdamW optimizer \cite{Loshchilov2019-Adamw} with weight decay of $2\times 10^{-5}$ and learning rate $3\times 10^{-4}$. The learning rate was halved upon validation loss plateau until it reaches a minimum of $10^{-5}$. A single Nvidia Tesla V100 GPU is used for training.


\subsection{Grain boundary structure dataset}

In this work we investigate a representative high-angle, high-energy $\Sigma5(100)$ twist boundary in refractory metal tungsten (W). 
Tungsten was chosen as a representative body-centered cubic (BCC) material and it is also a leading candidate for plasma-facing material (PFM) in fusion energy devices~\cite{marian_2017}. 
Tungsten GBs are known to be prone to brittle fracture~\cite{talignani_2022}, which motivates their investigation in this work. 
The twist boundary was obtained by a \SI{36.87}{\degree} rotation of the $(100)$-oriented W grains around a common $[001]$ axis with the GB plane perpendicular to the axis of rotation. 
The interaction between the tungsten atoms were modeled using an embedded-atom method (EAM) potential by Zhou, et al.~\cite{zhou_2004} 
GB structure search was performed using GRand canonical Interface Predictor (GRIP) \cite{chen_2024}. 
During the structure search, the tool systematically explores all the available GB microscopic degrees of freedom which include relative grain translations and different atomic densities at the boundary core by deleting and inserting a certain number of GB atoms to minimize the GB energy. 
The GB energy, $E_{\mathrm{gb}}$, is calculated as:
\begin{equation}
    E_{\mathrm{gb}} = \left( E^{\mathrm{gb}}_{\mathrm{total}} - N^{\mathrm{gb}}_{\mathrm{total}} E^{\mathrm{bulk}}_{\mathrm{coh}}\right)/{A^{\mathrm{gb}}_{\mathrm{plane}}},
    \label{eq:Egb}
\end{equation}
where $E^{\mathrm{gb}}_{\mathrm{total}}$ is the total energy of the region containing the GB, $E^{\mathrm{bulk}}_{\mathrm{coh}}$ is the energy per atom in the perfect W BCC crystal, and $A^{\mathrm{gb}}_{\mathrm{plane}}$ is the GB area.
The total number of nonequivalent atomic densities that should be sampled is given by the number of atoms in one bulk atomic plane parallel to the boundary because a complete removal of one plane of atoms coupled with the relevant translation is guaranteed to return the boundary to an identical state. 
As a result, this number of GB atoms, $[n]$, is measured as a fraction of the total number of atoms in one bulk atomic plane. 
If $N_{\mathrm{total}}$ is the total number of atoms in the bulk crystal and $N^{\mathrm{bulk}}_{\mathrm{plane}}$ is the total number of atoms in one bulk atomic plane, the fraction of GB atoms is 
\begin{equation}
    [n] = \left({N_{\mathrm{total}} \mod N^{\mathrm{bulk}}_{\mathrm{plane}}}\right)/{N^{\mathrm{bulk}}_{\mathrm{plane}}} \in [0, 1)
    \label{eq:n}
\end{equation}

The results of GB search with GRIP are summarized in Fig.~\ref{fig:GB-generation-conditional}. 
Each blue point corresponds to $E_{\mathrm{gb}}$ for a given atom fraction $[n]$. 
Here we generated 300 distinct structures with different energy and $[n]$. 
The search reveals two minima at $[n]=1/5$ and $[n]=2/5$ which correspond to the ground state and the metastable phase of this boundary, respectively (Fig.~\ref{fig:upfront-DPM-comparison}c). 
A previous study~\cite{frolov_2018_grain} demonstrated first-order structural phase transitions between these two GB phases at finite temperature when point defects were inserted in the ground-state (Fig.~\ref{fig:upfront-DPM-comparison}c). 
All other structures generated by GRIP have higher energy. 
Some of them are distorted versions of the two phases containing point defects, while other boundaries have completely different structures. 
While they are all mechanically stable at \SI{0}{\kelvin}, at finite temperature they are expected to evolve toward the two GB phases shown in Fig.~\ref{fig:upfront-DPM-comparison}c.

\paragraph{Metrics of generated structures.}
To quantitatively characterize the generated structures, we use the Steinhardt order parameter \cite{Steinhardt1983PRB} as the atomic descriptors, as well as radial distribution functions. 
For GBs, we focus on the energy $E_{\mathrm{gb}}$, as the goal is a generative model that can learn the complex GB atomic environments from a training set and efficiently predict new GB candidates that will be relaxed by energy minimization,
omitting the need for costly global optimization. 
This final minimization step is always necessary as generation of the exact ground state is unrealistic.

\section{Experiments}

\subsection{Simple ordered crystalline structures}
\begin{figure}[!ht]
    \centering
    \includegraphics[width=1.0\linewidth]{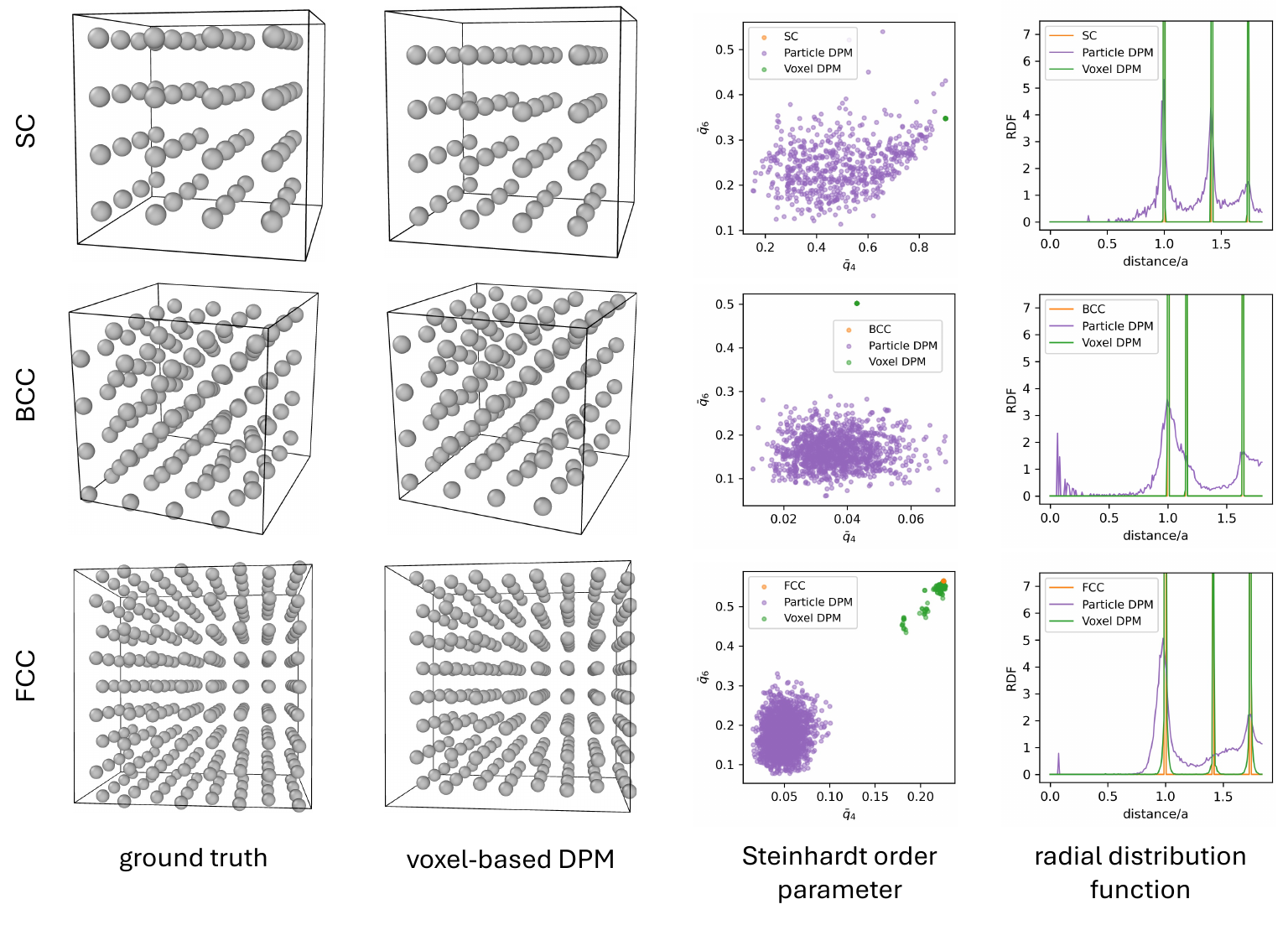}
    \caption{Ideal and generated structure for SC, BCC and FCC phases, and the corresponding Steinhardt order parameters and radial distribution functions averaged over ten $4^3$ cells.}
    \label{fig:crystalline-gt-pd}
\end{figure}

Separate particle- and voxel-based DPM models were trained for each of simple cubic (SC), FCC and BCC phases with one ideal $4\times 4 \times 4$ conventional cell as the training set. $L$ was chosen to partition the supercell into $24^3$, $20^3$ and $24^3$ voxels for SC, BCC and FCC, respectively.
Figure \ref{fig:crystalline-gt-pd} provides the ground truth and grand canonical generated (without relaxation) crystal structures for SC, BCC and FCC phases. The Steinhardt order parameters and radial distributions both show that the generated structures are very close to the reference structure.  In contrast, the particle-based baseline DPM produce structures that deviate significantly (purple points and curves).

\subsection{Conditional grain boundary generation by in-painting}\label{sec:inpainting}

To develop the grand canonical diffusion model for GB, we first smeared the W atoms using cube size $L=0.834$ \AA, resulting in a $12\times12\times72$ voxel array per structure. Then we trained the voxel-space diffuser, a 3D U-Net 
consisting of 2 levels and 3 residual convolutional blocks per level. The training lasts 30,000 steps with a batch size of 4.

\begin{figure}[!ht]
    \centering
    \includegraphics[width=0.7\linewidth]{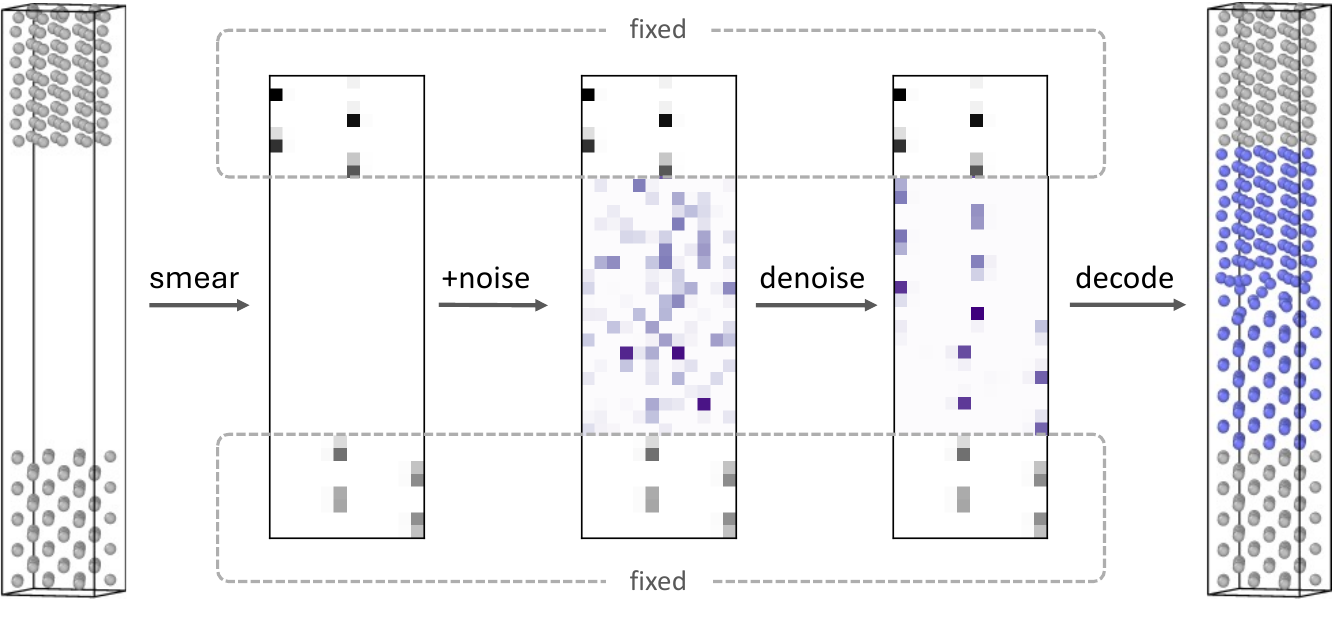}
    \caption{Conditional grain boundary generation via in-painting.}
    \label{fig:GB_inference}
\end{figure}

During inference, we adopt a conditional generation strategy by in-painting. 
Concretely, the initial input to the diffusion model is a voxel array encoded from perfect W BCC crystal grains on both ends of $z$ direction, each taking $12\times12\times18$ voxels and kept fixed during diffusion (Fig.~\ref{fig:GB_inference}). The middle of the array is subject to diffusion 
through the simulated annealing process to generate voxels in the GB region. Finally, the whole array is decoded into particles as described previously. 

\begin{figure}[!ht]
    \centering
    \includegraphics[width=0.99\linewidth]{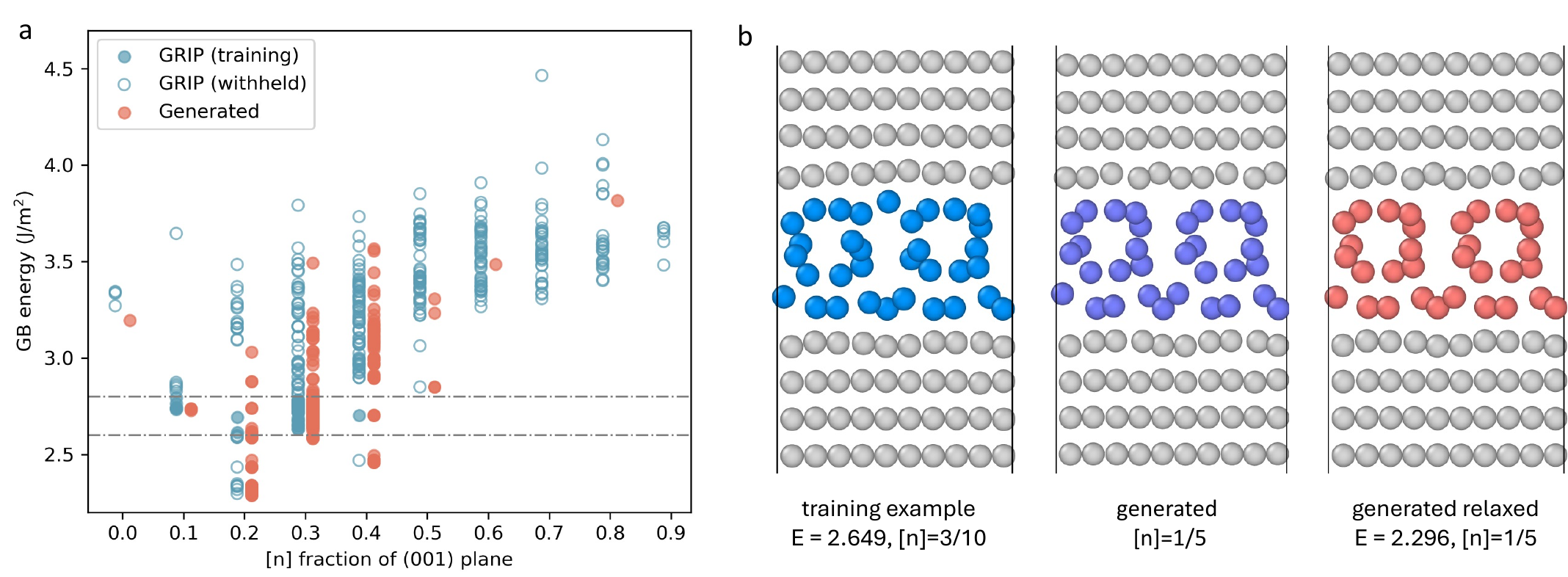}
    \caption{Grain boundary energy after relaxation and exemplar structures. (a) GB energy is plotted versus the number of atoms $[n]$ in the boundary core. GB structure search using GRIP tool (blue) and the developed diffusion model (red) both predict the ground state at $[n]=1/5$ and the second metastable phase at $[n]=2/5$ shown in Fig.~\ref{fig:upfront-DPM-comparison}c. Diffusion model is trained with GRIP generated structures with $E_{\mathrm{gb}}$ in the range of $ \SI{2.6}{\joule\per\meter\squared} \le E_{\mathrm{gb}} \le \SI{2.8}{\joule\per\meter\squared}$. (b) An exemplary structure used in training, the DPM generated structure and predicted relaxed ground state structure. }
    \label{fig:GB-generation-conditional}
\end{figure}

Trained on the full GB dataset, our model can successfully generate the low energy states (details in Appendix Fig.~\ref{fig:GB-generation-conditional_all_training}).
To show that it can not only memorize but also predict unseen structures, 
we built a training set consisting of GB structures with $\SI{2.6}{\joule\per\meter\squared} \le E_{\mathrm{gb}} \le \SI{2.8}{\joule\per\meter\squared}$, shown in Fig.~\ref{fig:GB-generation-conditional} as solid blue circles within the two dashed lines. The remaining data points are not used for training, depicted as blue open circles. 
Candidate GB structures were generated by the trained diffusion model and relaxed
with conjugate gradient minimization using LAMMPS~\cite{thompson_2022}. The convergence criteria are \SI{e-5} for relative energy 
and \SI{e-5}{\electronvolt\per\angstrom} for forces, with a maximum of \SI{e5} evaluations.
Note that since the model is grand canonical, it automatically generates candidate structures with different $[n] \in [0,1)$.

The results of the GB structure prediction 
are shown as red dots in Fig.~\ref{fig:GB-generation-conditional}a, next to the data points generated by GRIP. 
Remarkably, the diffusion model search predicted both the ground state at $[n]=1/5$ as well as the metastable GB phase $[n]=2/5$ (see Fig.~\ref{fig:upfront-DPM-comparison}c), which were excluded from the training set. Moreover, within the entire interval of the GB atomic densities, the distribution of generated $E_{\mathrm{gb}}$ for each fraction $[n]$ concentrates near the lowest energy configurations, illustrating that the generative model produces better GB candidates than those optimized  
by GRIP. Figs.~\ref{fig:GB-generation-conditional}b illustrates examples of a high energy, metastable GB structure from the training set with $E_{\mathrm{gb}}=\SI{2.649}{\joule\per\meter\squared}$ and $[n]=3/10$, a GB candidate structure as generated by the diffusion model, and the ground state predicted by the search with the diffusion model. While the metastable structure and the ground state are not identical, the diffusion model learns from the metastable configurations enough to generate a realistic GB structure closely resembling the ground state, making the structure prediction possible.

\section{Discussions}

In conclusion, we presented an alternative diffusion model with a voxel representation with continuously adjustable number of particles. Structures were sampled using simulated annealing in the space of voxel values, whereby creation and annihilation of particles occurs in a grand canonical ensemble, rather than moving a fixed number of particles in real coordinate space. Our method circumvents the challenge of global optimization of atomic structures directly. Instead, particles can be added or deleted continuously. The grand canonical diffusion model was tested successfully on crystalline phases including simple cubic, face centered cubic, and body centered cubic, 
as well as grain boundary structures in tungsten. Ground state GB structures withheld from the model can be discovered, suggesting that the grand canonical diffusion model was able to learn generalizable structural features with both order and disorder. Future work will explore application of our method for materials design and generative modeling of different types of grain boundaries, other defects such as dislocations, and bulk phases in alloys.

\paragraph{Limitations.}\label{sec:limitation} The presented method should be considered a first step towards generative models of crystalline or crystalline + defect structures in a grand canonical setting. There are a few notable technical limitations. (1) The adopted CNN architecture is only applicable to orthorhombic unit cells, and the lattice constants have to be integer multiples of the voxel linear dimension $L$ chosen at training time. (2) Since the data representations are voxel values rather than particle coordinates, conditional generation is less straightforward than in a particle representation. Usually a guiding likelihood $P(y| \boldsymbol{r})$ assumes known particle coordinates $\boldsymbol{r}$, and the noising voxel configuration $V_t$ needs to be decoded first to coordinates $\boldsymbol{r}_t= D(V_t)$, which is not well defined for $t>0$.  (3) Translational equivariance is valid only at integer multiples of $L$ but rather enforced through data augmentation during training. Rotational equivariance is not attempted in this work and can likewise be introduced approximately. (4) The current method takes  many voxels to represent each particle and has a big memory footprint. Future work is needed to resolve or mitigate such issues, such as adoption of more flexible score model architectures that allows different unit cell shapes and dimensions, vision transformers for conditional generation, equivariant constructs, and latent diffusion methods that are more memory efficient. The developed method can be used in areas such as materials design and grain boundary engineering. Any negative societal impact associated with those applications can be applied to our method.

\section*{Acknowledgements}
This work was performed under the auspices of the U.S. Department of Energy by Lawrence Livermore National Laboratory under Contract DE-AC52-07NA27344. This work was funded by the Laboratory Directed Research and Development (LDRD) Program at LLNL under project tracking code 22-ERD-016 and 23-SI-006. Computing support for this work came from the Lawrence Livermore National Laboratory institutional computing facility. TF was supported by the U.S. DOE, Office of Science under an Office of Fusion Energy Sciences Early Career Award. Computing support for this work came from the Lawrence Livermore National Laboratory Institutional Computing Grand Challenge program.

\printbibliography


\newpage
\appendix
\renewcommand\thefigure{S.\arabic{figure}}
\setcounter{figure}{0}
\section*{Appendix / supplemental material}

\begin{figure}[!ht]
    \centering
    \includegraphics[width=0.99\linewidth]{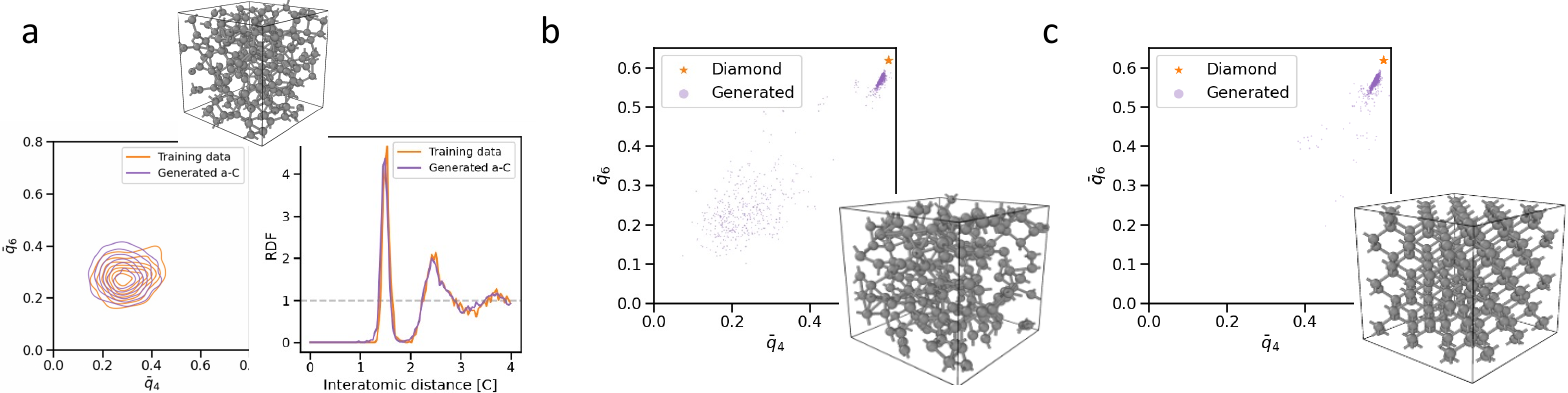}
    \caption{Carbon structure generation by simulated annealing using pretrained particle-based diffusion model from Ref.~\cite{Kwon2023-Spectroscopy}. Amorphous carbon (a), and diamond starting from diamond + larger (b) or small (c) perturbations. Amorphous structures have been successfully generated with random initial positions and real-space simulated annealing. Diamond can only be generated starting with slightly perturbed diamond, while initialization with large perturbations or random positions lead to significant amount of defects.}
    \label{fig:carbon}
\end{figure}

\begin{figure}[!ht]
    \centering
    \includegraphics[width=0.6\linewidth]{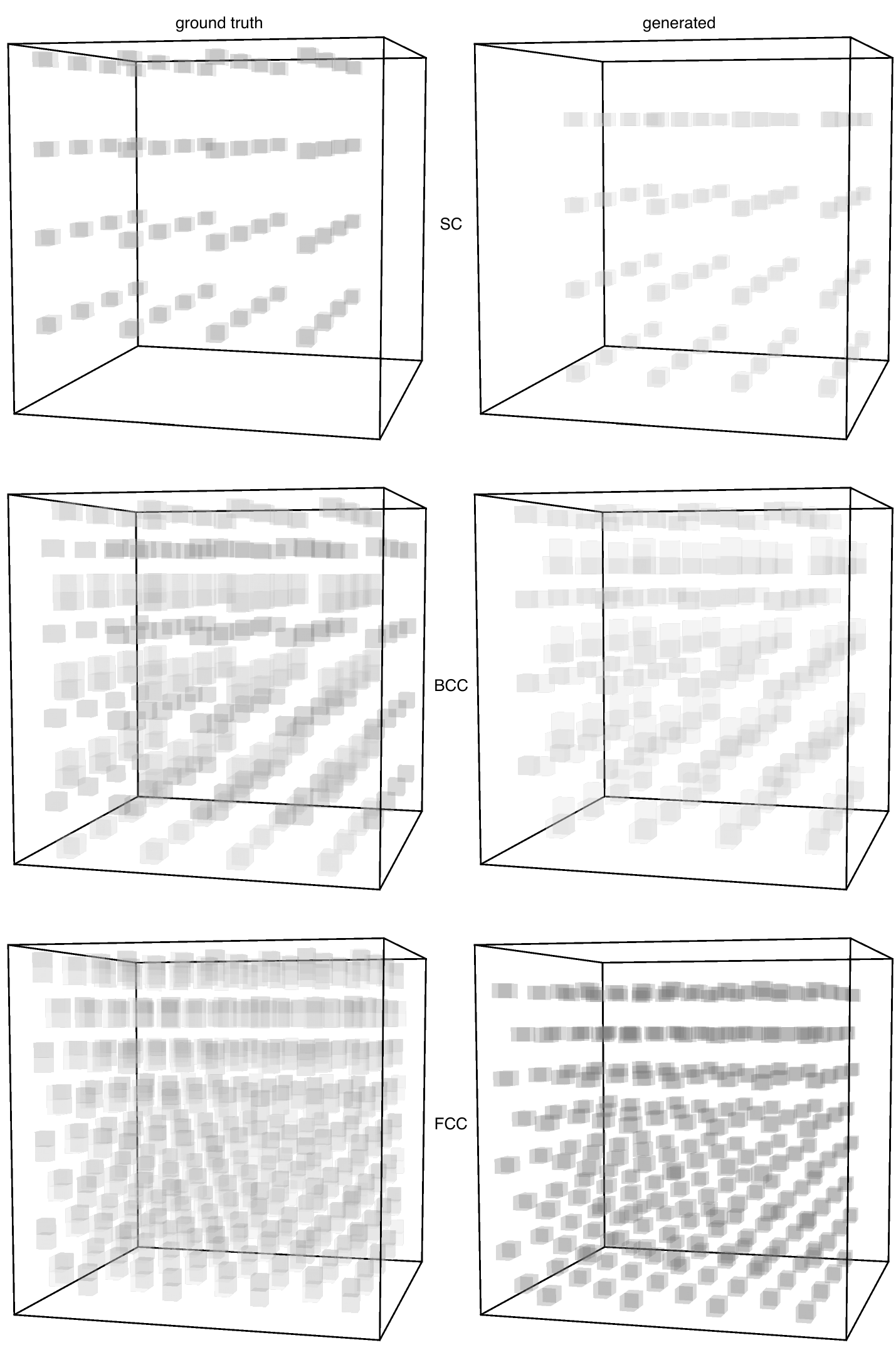}
    \caption{Voxel representation of crystalline phases: simple cubic (top row), BCC (middle) and FCC (bottom row). Left column: Encoded voxel arrays of  ideal structures. Right: generated configurations by grand canonical voxel-based DPMs separately trained on crystalline structure.}
    \label{fig:voxel-gt-pd}
\end{figure}

\paragraph{Details about particle-based GNN score model.}
The GNN score model $s_{\theta}(\mathbf{r}_t, t)$ is trained with dataset consisted of crystalline structures containing atomic coordinates $\mathbf{r}_0$ and unit cell dimensions $\mathbf{A}$. The score is learned via the denoising score matching loss
\begin{equation}
    \begin{aligned}
        \mathbf{r}_t 
            &= \alpha_t \mathbf{r}_0 + \sigma_t \epsilon, ~ \epsilon \sim \mathcal{N}(0,\mathbf{I}), \\
        \mathcal{L}_\text{DSM} 
            &= \mathbb{E}_{\mathbf{r}_0, \epsilon, t} \left[ 
                \Vert s_{\theta}(\mathbf{r}_t , t) \sigma_t + \epsilon \Vert^2
            \right],
    \end{aligned}
\end{equation}
where $\mathbf{A}$ is implicitly accounted for by the atomic graph representation. We apply the variance-exploding noise schedule such that $\alpha_t$ is fixed to 1 and $\sigma_t$ linearly increases from 0 (at $t=0$) to some maximum value $k$ (at $t=1$), which is set to 0.4 in this work. Under this variance-exploding scheme, we observe that the perturbed atomic coordinates $\mathbf{R}_t$ approach a random uniform distribution. Therefore, this enables one to start the denoising trajectory from a random structure sampled from a uniform distribution over the coordinate space.    

The GNN architecture consists of three components: an Encoder, a Processor, and a Decoder. The input atomic graph is denoted as $(V, E)$, where $V$ is a set of node features $\{\mathbf{h}_i\}$ for node/atom $i$, and $E$ is a set of edge/bond features $\{\mathbf{e}_{ij}\}$ from node $i$ to node $j$. The set of edges $E$ is determined by a cutoff distance $r_c$, within which a pair of nodes form an edge connection.

The Encoder is responsible for embedding node and edge features for the subsequent Processor operation. The node features are embedded as
\begin{equation}
    \mathbf{h}_i \gets M_H(\text{onehot}(\mathbf{z}_i)) + M_T(\text{GaussianRFF}(t)),
\end{equation}
where $M_H$ and $M_T$ are multilayer perceptrons (MLPs), and GaussianRFF is a shortname for the Gaussian Fourtier features \cite{tancik2020fourier}. Note that the atom type information $\mathbf{z}$ is optional in this work because we only focus on single-element systems in this paper. Regardless, we define the Encoder as above so that it can accommodate multi-element systems. The edge features $\mathbf{e}_{ij}$ are embedded as 
\begin{equation}
    \mathbf{e}_{ij} \gets M_E(\mathbf{r}_{ij} \oplus \bar{r}_{ij}),
\end{equation}
where $M_E$ is an MLP, $\mathbf{r}_{ij}$ are edge vectors computed as $\mathbf{r}_j - \mathbf{r}_i$, $\bar{r}_{ij}$ are edge lengths, and $\oplus$ denotes concatenation. $M_H$, $M_E$, and $M_T$ consist of two dense linear layers, with SiLU activation function after the first layer and layer normalization after the second.

The Processor is responsible for graph message-passing (convolutions). We used the Processor from MeshGraphNets \cite{Pfaff2020}, which we refer to for further details of the model. Note that the original work considers two sets of edges (mesh-space edges and world-space edges), whereas there is only one in our work.

The Decoder is responsible for mapping node features to score outputs. We use a simple MLP $M_O$ as the Decoder:
\begin{equation}
    \hat{y}_i \gets M_O(\mathbf{h}_i).
\end{equation}
$M_O$ has the same architecture as that of $M_H$, $M_E$, and $M_T$, except that there is no layer normalization at the end.

\begin{figure}[!ht]
    \centering
    \includegraphics[width=0.99\linewidth]{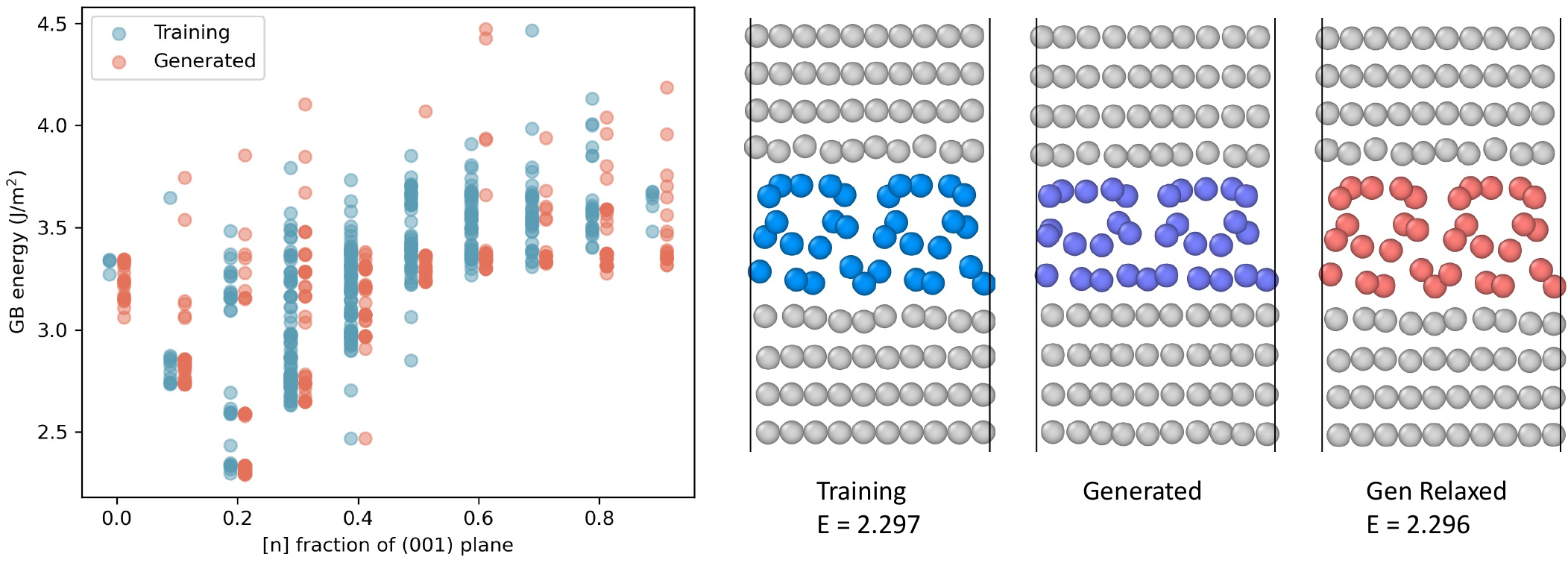}
    \caption{Same as Fig.~\ref{fig:GB-generation-conditional} but using all ground truth structures without holdout in training the grand canonical DPM. Grain boundary energy after relaxation (left) and exemplar structures (right).} 
    \label{fig:GB-generation-conditional_all_training}
\end{figure}

\end{document}